\documentclass[aip,jcp,preprint,floatfix,amsmath,amsfonts,14pt]{revtex4-1}
\usepackage[top=2cm,bottom=2cm,left=2.5cm,right=2.5cm]{geometry}

\usepackage{setspace}

\usepackage{graphicx}
\usepackage{amsmath}
\usepackage[utf8]{inputenc}
\usepackage{epstopdf}
\usepackage{color}
\usepackage{bbold}

\usepackage[normalem]{ulem}

\begin{document}

\title{Calculation of the static and dynamical correlation energy of pseudo-one-dimensional beryllium systems via a many-body expansion}
\author{D. Koch}
\author{E. Fertitta}
\author{B. Paulus}
\affiliation{Institut f\"ur Chemie und Biochemie - Takustr. 3, 14195 Berlin,\\ Freie Universit\"at Berlin, Germany}

\date\today

\begin{abstract}
Low-dimensional beryllium systems constitute interesting case studies for the test of correlation methods because of the importance of both static and dynamical  correlation in the formation of the bond. Aiming to describe the whole dissociation curve of extended Be systems we chose to apply the method of increments (MoI) in its multireference (MR) formalism. However, in order to do so an insight into the wave function was necessary. Therefore we started by focusing on the description of small Be chains via standard quantum chemical methods and gave a brief analysis of the main characteristics of their wave functions. We then applied the MoI to larger beryllium systems, starting from the Be$_6$ ring. First, the complete active space formalism (CAS-MoI) was employed and the results were used as reference for local MR calculations of the whole dissociation curve. Despite this approach is well established for the calculation of systems with limited multireference character, its application to the description of whole dissociation curves still requires further testing. After discussing the role of the basis set, the method was finally applied to larger rings and extrapolated to an infinite chain.
\end{abstract}

\maketitle

\vspace{1cm}
\section{Introduction}

Despite the tremendous progress made by \textit{ab initio} quantum chemical methods and algorithms in the past few decades enhanced by the constantly increasing computer power, a proper description of electronic correlation in extended and periodic systems still constitutes in many cases an unfeasible problem for standard wave function methods. Density functional theory (DFT)\cite{Dreizler1990,Eschrig1996,Kohn1965} generally allows to achieve reliable results for systems which could not be treated with single-reference post-Hartree-Fock methods because of their unfavourable scaling, but it fails in systems with strong static correlation which requires a multiconfigurational (MC) description. Furthermore, with DFT being not systematically improvable, wave function methods are always preferable when feasible. The need for novel approaches, which might deal with such strongly correlated and large systems, led to the development of a variety of approximations based on different approaches such as local methods\cite{Pulay1983,Pulay1986,Saebo1993,Stollhoff1991,Pardon1995,Kitaura1999,Schutz2000,Schutz2001,Schutz2002,Schutz2002a,Pisani2008,Pisani2012}, tensor product states\cite{Chan2009,Marti2010,Schollwock2011,Wouters-2014e,Murg2015,Szalay-2015} and/or stochastic approaches\cite{Scuseria2008,Booth2009,Spencer2012,Morales2012,Cleland2012,Shepherd2012,Kolodrubetz2012,Petruzielo2012,Roggero2013,Willow2014}.\\
In the framework of local correlation methods, approaches based on M\o ller-Plesset perturbation theory (MP)\cite{Pulay1983,Pulay1986,Pisani2008,Pisani2012} and coupled cluster theory (CC)\cite{Stollhoff1991,Schutz2000,Schutz2001,Schutz2002,Schutz2002a,Schutz2014,Schwilk2015} constitute a powerful and effective alternative to DFT, but once again their single-reference formalism is not suitable for dealing with strongly correlated electrons. Among other local approaches, also the method of increments (MoI)\cite{Stoll1992a,Stoll1992b,Stoll2009,Stoll2010,Paulus2003,Paulus2006,Voloshina2007,Schmitt2009,Muller2011,Voloshina2012,Paulus2003a,Alsheimer2004} offers a powerful tool for calculating the correlation energy of extended and periodic systems. This method, which is based on a many-body expansion of the correlation energy in terms of localized molecular orbitals, can be used in different formalisms along with any size-extensive correlation method. This flexibility of the MoI in the past allowed for a multireference (MR) formalism to accurately calculate the ground state energy of different bulk metals in their equilibrium structure.\cite{Voloshina2014}\\
Moreover, in a recent work\cite{Fertitta2015} Fertitta, Paulus \textit{et al.} applied a complete active space self consistent field (CAS-SCF) formalism of the MoI (CAS-MoI) to calculate the dissociation curves of highly correlated pseudo-one dimensional systems, such as beryllium rings. By comparison with the benchmarks obtained with various methods, including the \textit{ab initio} density matrix renormalization group (DMRG)\cite{White1992,White1993,Chan2009,Marti2010,Schollwock2011,Wouters-2014e,Olivares-Amaya-2015,Murg2015,Szalay-2015}, we showed how accurate results can be obtained via MoI also for regions of the dissociation curve close to avoided crossings where the static correlation drastically increases. The Be ring system was chosen to model periodic one-dimensional arrangements in order to test the CAS-MoI. Indeed, by  exploiting the locality of the method we were able to calculate the correlation energies of a system as large as Be$_{90}$ and via extrapolation the value for the limit corresponding to the infinite chain could be evaluated. However, this preliminary work was carried out using a very poor one-particle description of our model system, that is with a minimal basis set $(8s,3p) \rightarrow [2s,1p]$, neglecting the effect of dynamical correlation and the influence of more diffuse basis functions. Therefore, in the present work we aim to apply this approach to investigate the behavior of the MoI in a more sophisticated formalism which allows to include this effects. We will therefore employ larger basis sets and apply multireference methods on top of CAS-MoI calculations and discuss the contributions to the total electron correlation yielded by the different approaches.\\
Before dealing with the method of increments treatment, however, we will focus on small Be clusters which can be described by means of standard methods. By doing so and analyzing the respective dissociation curves as well as the evolution of the wave functions, we will gain a better understanding of the problematics arising when dealing with large beryllium chains.\\
This paper is structured as follows: in section~\ref{sec_comp_det} we describe the details of the calculations and justify our choice of a proper basis set; in section~\ref{sec_small_cluster} we report the dissociation curves of small chains discussing the change in the nature of the bond with system size as well as the problems arising as the required active space grows with it; in section~\ref{sec_moi} the method of increments is presented in its different formalisms as applied in this work and the results obtained by applying it for Be$_6$ and larger clusters are reported and discussed in section~\ref{sec_results}; finally we draw our conclusions in section~\ref{sec_conclusion}.

\section{Computational details}\label{sec_comp_det}

\begin{figure}[h]
    \includegraphics[width=0.5\textwidth]{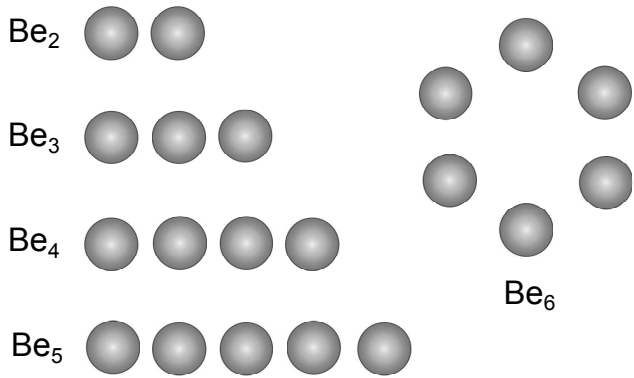}
    \caption{(Color online) Schematic depiction of the Be arrangements discussed in sections~\ref{sec_small_cluster} and \ref{be6_moi}, including Be$_n$ linear chains with $n$ up to 5 and a Be$_6$ ring with equidistant spacings between the atomic centers.}
\label{fig_1}
\end{figure}

\begin{figure}[h]
    \centering{    
    \includegraphics[width=0.5\textwidth]{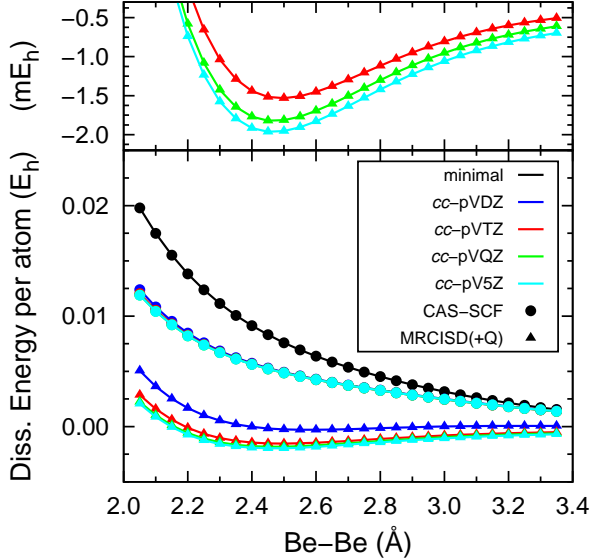}}
    \caption{(Color online) Dissociation curve of the beryllium dimer calculated using CAS(4,8) and a successive MRCISD(+Q). Different basis sets $cc$-pVXZ were employed as well as a minimal $(9s,4p) \rightarrow [2s,1p]$ basis set for comparison. The close-up in the upper panel highlights the difference among $cc$-pVTZ, $cc$-pVQZ and $cc$-pV5Z.}
\label{fig_2}
\end{figure}

All calculations, including Hartree-Fock (HF), post-HF, localization and MoI, were performed employing the quantum chemical program package MOLPRO\cite{MOLPRO}. The localizations were performed using the Foster-Boys\cite{Foster1960} procedure. All the different dissociation curves presented in this work are reported as a function of the Be-Be internuclear distance. Independently of the choice of our model, a linear chain or a ring, we always imposed the condition of equal nearest neighbour distances in the whole system. In the case of rings this means that the symmetry $D_{nh}$ is conserved all over the dissociation curve. Fig.~\ref{fig_1} schematically depicts the shapes of the Be systems dealt with in sections~\ref{sec_small_cluster} and \ref{be6_moi}. In order to allow an easier comparison among the different systems, we will always report energies per atom, whether we are dealing with dissociation energies or correlation energy contributions. It has to be underlined that no counterpoise correction was applied for the calculation of dissociation energies, since we aimed mainly to compare different methods rather than to achieve an accurate result for comparison with experiments.\\
The basis sets for the larger systems were selected after performing different tests on the beryllium dimer. These are shown in Fig.~\ref{fig_2}, where we compare CAS(4,8) and CAS(4,8)+MRCISD(+Q) (multireference configuration interaction singles and doubles with Davidson corrections) calculations using different basis sets of the Dunning's family\cite{Prascher2011} $cc$-pVXZ (with X = D, T, Q, 5) and a minimal $(9s,4p) \rightarrow [2s,1p]$ basis set derived from a contraction of $cc$-pVDZ. As can be seen, in order to achieve reasonable qualitative and quantitative results, the VTZ basis set can be considered reliable. The results obtained with the minimal basis set, on the other hand, deviate a lot from the CAS(4,8) obtained with the other basis sets. Nevertheless, we will keep track of the results achievable with this basis set, since this can be used to understand the qualitative structure of the wave function and allow to easily test the local method that we will employ for large beryllium rings.

\section{Electronic structure of small beryllium systems}\label{sec_small_cluster}

\subsection{Linear chains}

\begin{figure}[h]
    \centering{    
    \includegraphics[width=0.5\textwidth]{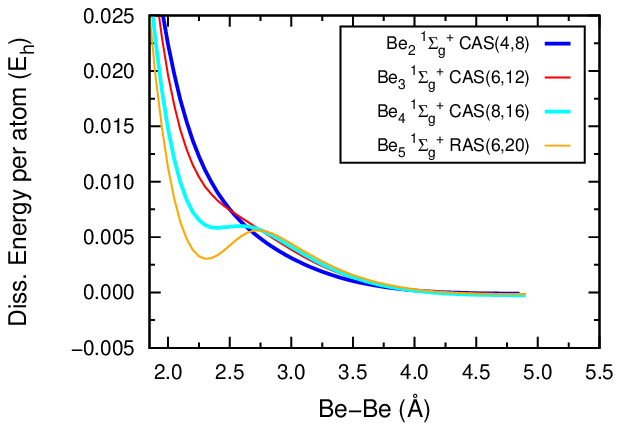}\\
    \includegraphics[width=0.5\textwidth]{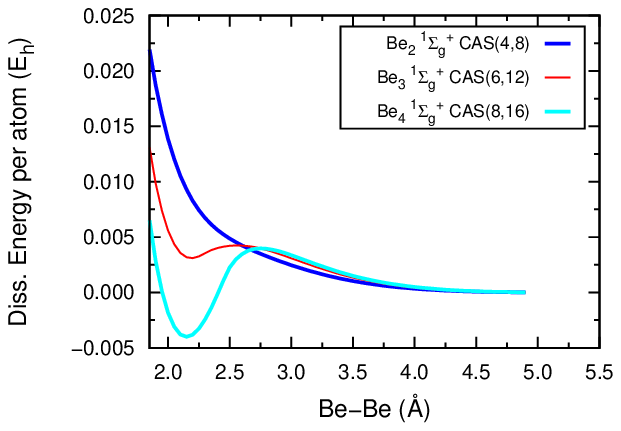}\\
\includegraphics[width=0.5\textwidth]{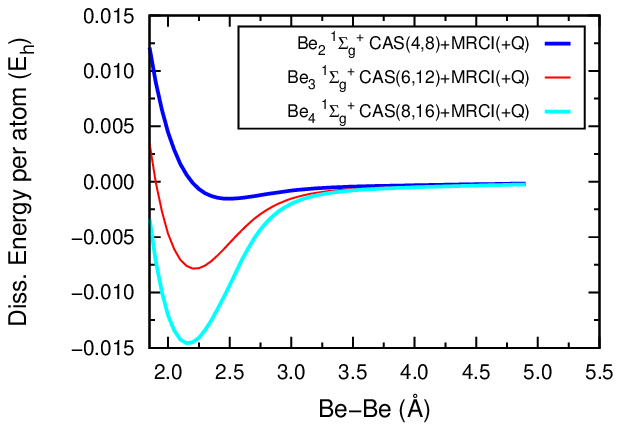}}
    \caption{(Color online) Dissociation curves of small linear beryllium chains calculated with different methods and basis sets. From the upper to the lower panel: CAS(2$n$,4$n$) with a minimal basis set; CAS(2$n$,4$n$) with $cc$-pVTZ; CAS(2$n$,4$n$)+MRCISD(+Q) with $cc$-pVTZ.}
\label{fig_4}
\end{figure}

As already stated, the interest towards these systems lies in the important role played by static correlation in the bonding, which becomes incredibly hard to correctly describe as system size increases. The bond of a beryllium dimer has been subject of many investigations\cite{Merritt2009,Patkowski2007,Koput2011,Sheng2013} showing how a multireference approach is crucial for achieving a quantitative description of the dissociation. Indeed, single-reference methods, such as truncated configuration interaction (CI) and coupled cluster, fail in such a task\cite{ElKhatib2014,Heaven2011}. This can be seen by comparing the dissociation energies obtained through different methods for different chain lengths which are reported in Table~\ref{tab_1}. In the case of Be$_2$, CI singles and doubles (CISD) underestimates the dissociation energy of one order of magnitude, while CC singles and doubles (CCSD) yields a repulsive dissociation curve. The inclusion of perturbative triples by CCSD(T) allows for achieving a minimum which, however, still strongly underestimates the dissociation energy. On the other hand, better results are obtained by means of multireference CI methods, such as MRCISD, MRCISD(+Q) and the averaged coupled pair functional (ACPF). In particular, the latter yields a dissociation energy in excellent agreement with full CI (FCI), namely -1.42~m$E_h$. However, the choice of the multiconfigurational reference is crucial. As expected and widely explained by Evangelisti \textit{et al.}~\cite{Helal2013,ElKhatib2014}, a full valence CAS(4,8) reference guaranties to achieve this goal. On the other hand, CAS(2,2) gives inconsistent results, but a CAS(4,4), which includes $\sigma$ orbitals only, allows to obtain reliable data if a multireference method is applied on top of it. This can be seen again by comparing the MRCI data in Table~\ref{tab_1}.\\
Analyzing the data for the longer chains, one can see how the dissociation energy per atom increases rapidly and independently of the method employed. This indicates a sudden change in the character of the bond which is dominated by dispersive forces in the dimer. CISD and CCSD still heavily underestimate the dissociation energy, while CCSD(T) seems to retrieve better values as the number of atoms increases. These numbers, however, cannot be fully trusted either,  since the $\mathcal T_1$ diagnostics yields values of 0.06 for Be$_3$ and longer chains, which is definitely higher than the recommended threshold of 0.02-0.025. This indicates a pronounced multiconfigurational nature of the wave function that seems to increase with system size. Moreover CCSD shows instabilities in the convergence procedure in the region of the minimum, underlining, once again, the necessity of employing a multireference approach.\\
As one can see, the MRCISD results obtained by using a CAS($2n$,$4n$) and a CAS($2n$,$2n$) references are in general in good agreement also when the Davidson corrections are included. Indeed, close to the minimum, the configurations with the largest weight involve only $\sigma$ orbitals and $\sigma\rightarrow\pi$ excitations become important at larger internuclear distances as shown in the supplementaty materials. In Fig.~\ref{fig_3} we show the weights of the two main configurations for Be$_n$ chains with $n=2,3,4,5$ as calculated with a minimal basis set. It is evident that as system size grows, the multireference character increases especially for short internuclear distances, where the two configurations have comparable weights.\\
In Fig.~\ref{fig_4} we compare the dissociation curves of beryllium chains of different lengths in their ground state $^1\Sigma_g^+$. Going from the upper to the lower panel, we report the results obtained at the CAS($2n$,$4n$) level with the minimal and VTZ basis set and the dissociation curves calculated with MRCISD(+Q) on the top of the CAS-SCF wave function with VTZ basis set. It has to be underlined that for Be$_5$ the complete active space (10,20) was already too large, so that we performed restricted active space (RAS)-SCF calculations using 6 active electrons in 20  orbitals. The accuracy of these results is discussed in the supplementary materials. We will start with the evaluation of the CAS-SCF results with the minimal basis set. As it can be seen, even at this low level of approximation the character of the bond changes drastically when going from the dimer to longer chains. Indeed, from a purely repulsive curve for Be$_2$ and Be$_3$, a minimum at around {2.3~\AA} appears for Be$_4$ and Be$_5$ as well as a barrier at {2.7~\AA}. Despite the minimum is energetically unfavorable with respect to the dissociation limit, the dissociation energy per atom gets significantly lower as system size increases. The situation is more pronounced when the VTZ basis set is employed, since in this case already for Be$_3$ a minimum occurs. The main characteristics of the dissociation curve are unchanged despite a shift of the minimum and the barrier towards smaller internuclear distances, but the dissociation energy decreases and Be$_4$ is already weakly bound at this level of theory with a dissociation energy per atom in the order of 4~$mE_\mathrm{h}$. As already stated, this substantial change in the strength of the Be-Be bond from the dimer to longer chains indicates a change in the nature of the bond itself, which seems to evolve from purely dispersive to a covalent or metallic character which can be qualitatively described by introducing static correlation at the CAS-SCF level. Finally, when introducing dynamical  correlation, all clusters are bound even though the differences between Be$_2$ and the longer chains are evident. The dissociation energy increases in magnitude by roughly 10~$mE_\mathrm{h}$ with respect to the CAS-SCF calculations and the barrier disappears.\\
Keeping in mind that we are interested in understanding how the binding situation changes towards the thermodynamical limit, one should perform similar calculations on larger systems. However, as already stated, even for Be$_5$ we had to employ a RAS reference wave function instead of the full valence CAS-SCF and MRCI on top of this was not feasible.

\begin{table*}
\centering
\caption{Dissociation energy  per atom of small beryllium chains calculated at different level of theory with a $cc$-pVTZ basis set in the respective minima, which are reported in the supplementary materials. The values are reported in m$E_h$. The symbol ``!'' indicates a repulsive dissociation curve. Dissociation energies are calculated with respect to the energy at internuclear distance Be-Be=10.0\AA.}\label{tab_1}
\begin{tabular}{lp{0.2cm}cccp{0.2cm}ccp{0.2cm}ccc}
          &&&&&& \multicolumn{2}{c}{CAS(2$n$,2$n$)} && \multicolumn{3}{c}{CAS(2$n$,4$n$) } \\
          \cline{7-8}\cline{10-12}

	  && CISD & CCSD & CCSD(T) &&  MRCISD & MRCISD(+Q) && MRCISD   &  MRCISD(+Q) & ACPF \\
\hline                                                                                                                                                  
Be$_2$     && -0.13&   !  & -0.90&& -1.37 &  -1.40         &&  -1.38 & -1.53   &  -1.45  \\
Be$_3$     && -2.24& -2.40& -7.43&& -7.68 &  -7.02         &&  -7.32 & -7.82   &  -7.66  \\
Be$_4$     && -6.35& -8.54& -15.5&& -14.7 &  -14.0         &&  -14.0 & -14.6   &  -14.4  \\
Be$_5$     && -10.6& -13.8& -21.0&& -19.5 &  -19.0         &&    --  & --      &     --    
\end{tabular}
\end{table*}

\begin{figure}[h]
    \centering{    
    \includegraphics[width=0.5\textwidth]{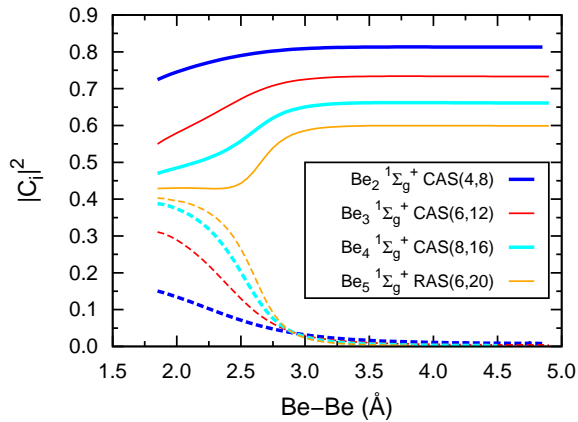}}
    \caption{(Color online) Weights of the two leading configurations for the dissociation of Be$_n$ chains (with $n=2,3,4,5$) calculated at the CAS(2$n$,4$n$) level with a minimal $(9s,4p) \rightarrow [2s,1p]$ basis set. Full and dashed lines indicate the most important and second most important configurations, respectively.}
\label{fig_3}
\end{figure}   

\subsection{Ring shaped systems}

\begin{figure}[h]
\includegraphics[width=0.5\textwidth]{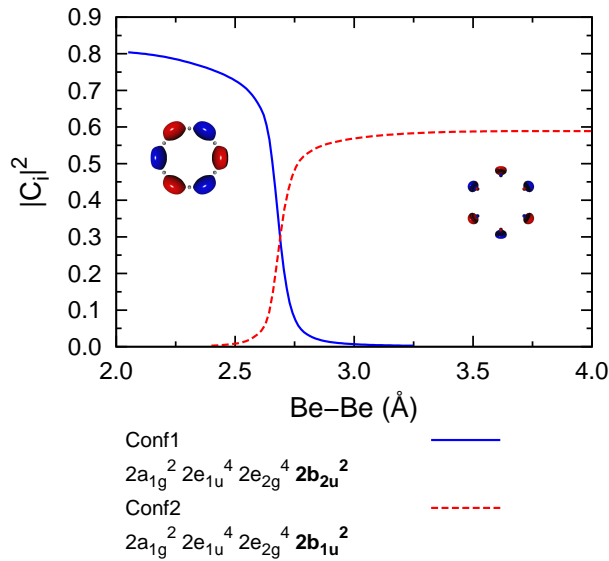}
\caption{(Color online) Weights of the two leading configurations (\textit{Conf1} and \textit{Conf2}) of the ground state of the Be$_6$ ring calculated with a RAS(4,24) calculation employing a minimal basis set. The valence occupations of \textit{Conf1} and \textit{Conf2} are given beneath in terms of the molecular orbitals in $D_{6h}$ symmetry. In the insets, depictions of the $2 b_\mathrm{2u}$ (left) and $2 b_\mathrm{1u}$ (right) orbitals are shown.}
\label{fig_5}
\end{figure}

So far we have discussed the electronic structure of linear beryllium chains which could be succesfully calculated up to Be$_4$ applying a CAS(2$n$,4$n$)+MRCI(+Q). As system size increases and we move towards the thermodynamical limit, also the use of CAS(2$n$,2$n$) or RAS-SCF is destined to be unfeasible and a local approach becomes a more (if not the only) efficient way to proceed. In order to do so, however, it becomes preferable to adopt different boundary conditions for our system. Indeed, by using a cyclic cluster one can fully exploit its rotational symmetry which allows one to reduce the number of individual local calculations that must be performed. Furthermore, the use of rings gives us the chance of moving naturally towards a description analogous to that offered by the Born von Karman boundary conditions which are used in periodic calculations.\\
Our investigation will focus first on the Be$_6$ ring system. As discussed in the previous works\cite{Fertitta2014,Fertitta2015} by Fertitta, Paulus \textit{et al.}, $p$-functions of Be rings do not only play an important role for a multiconfigurational treatment, but are also strongly affecting the Hartree-Fock wave function. Indeed, for the ground state ($^1A_{1g}$ in $D_{6h}$ symmetry) HF configuration, the character of the HOMO strongly varies along the dissociation curve. If we consider a minimal basis set, for large interatomic distances, it is an antibonding linear combination of $2s$ atomic orbitals, while for short interatomic distances the HOMO shows a pure $p$-character (see insets in Fig. \ref{fig_5}). As a consequence, its symmetry changes from $b_{1u}$ to $b_{2u}$. The ground state HF configurations arising from such a situation are labeled as \textit{Conf1} and \textit{Conf2} for short and large interatomic distances, respectively, and reported beneath Fig.~\ref{fig_5}. In this figure we also report the weights of these two configurations, as square of the corresponding CI coefficients, for the ground state of Be$_6$ ring  as obtained from a RAS(4,24) calculation. Once again, one can see that \textit{Conf1} is the configuration with the highest weight at short Be-Be distances while \textit{Conf2} is predominant at dissociation. At around 2.8~{\AA}, a crossing occurs and it is evident that a multireference treatment becomes necessary in this region.\\
The electronic structure of Be rings is very reminiscent of the one of linear chains, with a few differences. Indeed, since the point group is reduced from $D_{\infty h}$ to $D_{nh}$, the $\pi$ orbitals clearly lose their degeneracy and the ones lying on the plane of the molecule mix with the $\sigma$ orbitals. Since a clear distinction between $\sigma$ and $\pi$ orbitals cannot be done anymore, CAS(2$n$,2$n$) calculations are no longer possible. On the other hand, the HOMO and LUMO will have a pure $s$ or $p$ character because of these boundary conditions. As an indirect consequence, two distinct configurations will be dominant in different regions of the dissociation curve. This will play in our favour when applying the MoI. Moving forward in our discussion, we will then switch to this structure in order to model one-dimensional beryllium systems.

\section{The method of increments}\label{sec_moi}

\begin{figure}[h]
\includegraphics[width=0.5\textwidth]{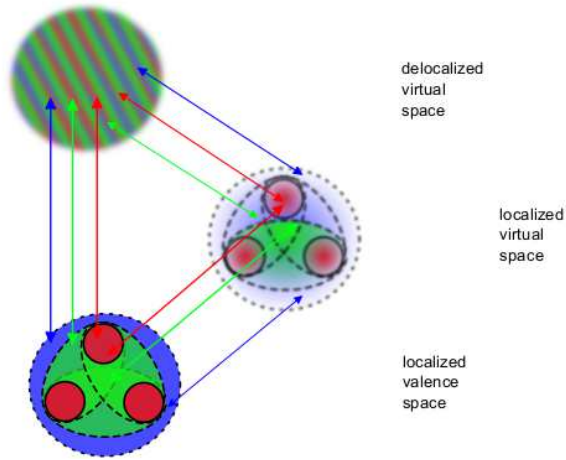}
\caption{(Color online) Schematic illustration of the method of increments. The areas of different color represent the one-body (red), two-body (green) and three-body (blue) fractions of the correlation energy in an assembly with three centers. Excitations into the different virtual spaces are depicted by arrows. In the single-reference case only occupied orbitals are localized, whether in the multireference formalism also properly chosen virtuals are localized. In this case the static correlation contribution is calculated using LOs only, while the excitations into the delocalized virtuals allow to evaluate the dynamical contribution.}\label{fig_6}
\end{figure}

Due to the short-range nature of electron correlation, many successful model Hamiltonians like Hubbard's yield surprisingly good predictions, just considering nearest neighbor interactions. A similar philosophy is adopted by quantum chemical local methods, which allow to retrieve a sometimes impressive amount of the correlation energy of extended systems, exploiting an expansion in terms of contributions from locally limited parts of the system. For this purpose, one has to perform first an unitary transformation of the molecular orbital basis in order to obtain localized orbitals (LOs) which will then be employed as a new orbital basis for the post-Hartree-Fock calculations.\\
Within the method of increments, correlation calculations are carried out with a properly defined set of orbitals allocated at specific centers to which we will refer as bodies. This allows to evaluate contributions to the total correlation energy, $E_\mathrm{corr}$, associated with different regions of the system which can be finally summed up.\\
Once the $N$ bodies in which the system has been split have been chosen, a first approximation for $E_\mathrm{corr}$ is given by the sum of all individual $N$ correlation energy contributions associated with each body:
\begin{equation}
E_\mathrm{corr}^\mathrm{(1)} =\sum_{\mathrm{i}}^\mathrm{N} \epsilon_\mathrm{i}
\label{moi1}
\end{equation}   
We will refer to these individual contributions $\epsilon_\mathrm{i}$ as one-body increments. At the one-body level, a significant fraction of $E_\mathrm{corr}$ can be retrieved if the bodies are chosen in a reasonable fashion. This is of course not enough to obtain highly precise predictions concerning chemical processes. However, by introducing contributions derived from higher order increments one can achieve such a goal.
These can be calculated by considering the correlation between two bodies, three bodies and so on. Therefore, sets of local orbitals at multiple bodies are included in correlation calculations leading to values $\epsilon_\mathrm{ij\dots z}$. By subtracting the values corresponding to the respective lower order increments one can calculate the required terms. As an example, for the two-body increment we have the expression:
\begin{equation}
\Delta \epsilon_\mathrm{ij} = \epsilon_\mathrm{ij} - (\epsilon_\mathrm{i} + \epsilon_\mathrm{j})
\label{moi2}
\end{equation}  
This procedure can be extended with more and more bodies taken into account, analogously subtracting all lower-order contributions, therefore the three-body increment can be expressed as:
\begin{equation}
\Delta \epsilon_\mathrm{ijk} = \epsilon_\mathrm{ijk} - (\Delta \epsilon_\mathrm{ij} + \Delta \epsilon_\mathrm{jk} + \Delta \epsilon_\mathrm{ik}) - (\epsilon_\mathrm{i} + \epsilon_\mathrm{j} + \epsilon_\mathrm{k})
\label{moi3}
\end{equation} 
Finally, the total correlation energy can be evaluated including all contributions: 
\begin{equation}
E_\mathrm{corr} =\sum_{\mathrm{i}} \epsilon_\mathrm{i} + \sum_{\mathrm{i<j}} \Delta \epsilon_\mathrm{ij} + \sum_{\mathrm{i<j<k}} \Delta \epsilon_\mathrm{ijk} + \cdots
\label{moi4}
\end{equation}    
Once again, since the electron correlation is short-ranged, it is generally safe to state that if the distance $r$ between the contributing bodies or the incremental order increases, the incremental contributions decrease:
\begin{equation}
|\Delta \epsilon_\mathrm{ij}| > |\Delta \epsilon_\mathrm{ik}| ~~\mathrm{for}~~ r_\mathrm{ij} < r_\mathrm{ik}
\label{moi5}
\end{equation}
\begin{equation}
    |\Delta \epsilon_\mathrm{ij}| > |\Delta \epsilon_\mathrm{ijk}| > |\Delta \epsilon_\mathrm{ijkl}|
\label{moi6}
\end{equation}
If the above relations are fulfilled and the expansion converges rather quickly, a reasonable truncation of the expression in Eq.~(\ref{moi4}) can be done and the method can be successfully applied. 
The advantage of such a procedure is obvious. Instead of performing one single expensive (or even unfeasible) calculations, our task will be rather limited to several smaller calculations.\\
So far we have been quite vague on the choice of LOs constituting the bodies as well as concerning which molecular orbitals are considered for the localization procedure. This is because there exists no absolute ``recipe'' for this, but it depends rather on the chemistry and physics of the system. Moreover, there are different formalisms of the MoI which depend on the extent of the static correlation. However, independently of the choice of the formalism employed, the equations described in this section are always valid.\\
Depending on the electron correlation method applied, different localization patterns are necessary. Within this work, three approaches were used along with the method of increments:
\begin{enumerate}
\item when using a single-reference method (CCSD(T)), only the occupied orbitals are localized. Excitations  are then allowed into the complete delocalized virtual space;
\item if a multiconfigurational method (CAS-SCF) has to be applied, both occupied and a properly chosen set of virtual orbitals has to be localized. These are then used in the MC-SCF procedure allowing orbital relaxation. It has to be underlined that only the LOs constituting the chosen body (or bodies) are optimized, since the others are kept frozen. This approach allows  to obtain the static part of the electron correlation;
\item a multireference treatment on top of the multiconfigurational calculation can be performed allowing for excitations from the localized active space into the delocalized virtual space to obtain the dynamical  contributions of the electron correlation. As MR approaches we employed MRCISD with (+Q) and without Davidson correction and ACPF.
\end{enumerate}
We will refer to these approaches as CCSD(T)-MoI, CAS-MoI, MRCISD-MoI, MRCISD(+Q)-MoI and ACPF-MoI, respectively.
A schematic depiction of the formalism is given in Fig.~\ref{fig_6}. This is calculated as the sum of the HF energy and all incremental contributions to the correlation energy up to a specific body order.

\section{Results}\label{sec_results}

\subsection{Be$_6$ Ring}\label{be6_moi}

\begin{figure*}[ht]
\includegraphics[width=0.5\textwidth]{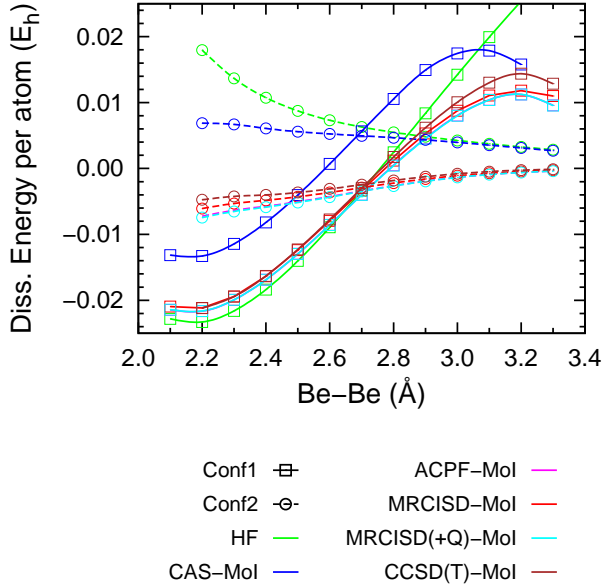}
\caption{(Color online) Dissociation curves of the Be$_6$ ring calculated with the method of increments at different levels of theory and using $cc$-pVDZ basis set. Two starting configurations \textit{Conf1} and \textit{Conf2} were employed, leading to a crossing.} 
\label{fig_7}
\end{figure*}

In this section we will discuss the results obtained for the Be$_6$ ring using different formalisms of the method of increments and we will briefly analyze the differences among the different approaches, focusing on the effect of static and dynamical correlation on the overall dissociation energy, on the position of the crossing and on the individual increments. This will serve to gain insights regarding the application of the method before moving to larger rings.\\
In order to apply the MoI, we will employ as reference the two configurations \textit{Conf1} and \textit{Conf2}, dominating in the minimum regime and towards dissociation, respectively. By performing a localization procedure on both configurations, two different sets of localized orbitals emerge, where the LOs obtained by unitary transformation of the orbitals from \textit{Conf1} are bond-centered, while those from \textit{Conf2} are centered near the positions of the nuclei, indicating the bonding and non-bonding character of the two configurations.\\
Let us start by describing the results obtained with the $cc$-pVDZ as an illustrative example, before applying the same procedure with larger basis sets. Fig.~\ref{fig_7} shows the dissociation curves calculated with HF, CAS-, MRCISD-, MRCISD(+Q)-, ACPF- and CCSD(T)-MoI using \textit{Conf1} and \textit{Conf2} as starting configurations. At each level of theory, the two curves obtained by these references cross: the energies obtained starting from \textit{Conf1} are lower than those for \textit{Conf2} for short interatomic distances and viceversa at larger bond lengths. We had already highlighted the inability of the approach to describe this avoided crossing\cite{Fertitta2015}, but by employing high accuracy benchmarks we concluded that the error in the energy was very small in this regime, too.\\
As for small linear chains, the minimum of the dissociation curve occurs at around 2.20~{\AA} for all methods. As one can see, the HF minimum of the dissociation curve is lower in energy than the ones obtained by post-HF-MoI calculations. This should not be surprising because of the large correlation contributions necessary to correctly describe the dissociation limit. By including static correlation by CAS-MoI, a dissociation curve reminescent of the ones obtained for Be chains at the CAS-SCF level is obtained: the system is weakly bound and a small barrier appears as a consequence of the repulsive dissociation curve yielded by \textit{Conf2}. The inclusion of static electron correlation has a large impact on the position of the crossing which is shifted to around 2.70~{\AA} in contrast to 2.90~{\AA} in Hartree-Fock.\\
The inclusion of dynamical correlation is achieved by different methods which yield similar results independently from the choice of a single-reference or a multireference approach. MRCISD(+Q)- and ACPF-MoI dissociation energies are almost identical for all distance regimes and they are only slightly lower than the ones obtained with CCSD(T)- and MRCISD-MoI, with a difference of just 3~m$E_h$ in the minimum.\\

\begin{figure*}[ht]
\includegraphics[width=\textwidth]{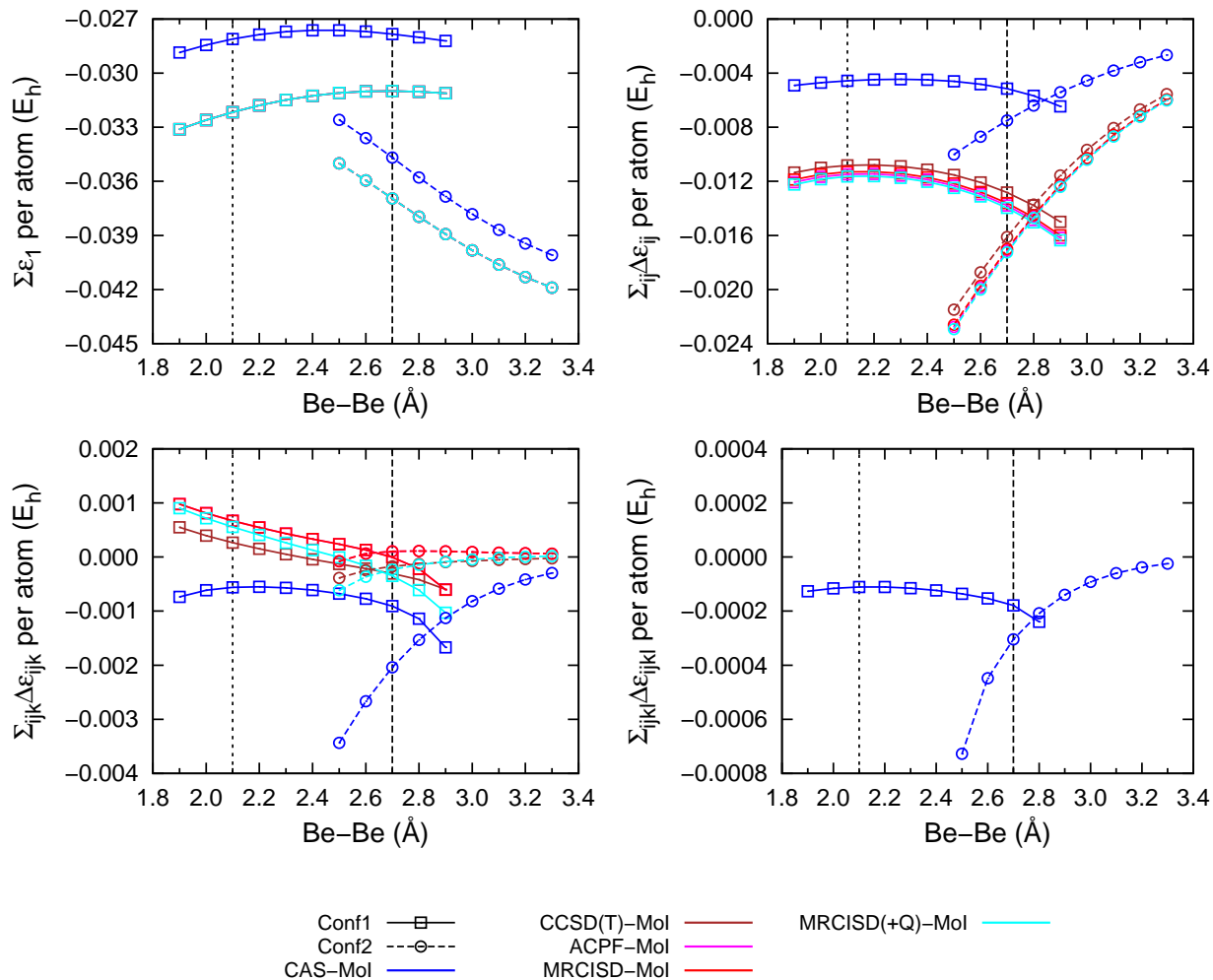}
\caption{(Color online) Correlation energy contributions of the Be$_6$ ring calculated with the method of increments at different incremental orders. Two starting configurations \textit{Conf1} and \textit{Conf2} with $cc$-pVDZ basis set were employed. The minimum as well as the crossing regime are indicated by a dotted and a dashed vertical line, respectively.}
\label{fig_8}
\end{figure*}

\begin{figure*}[ht]
\includegraphics[width=\textwidth]{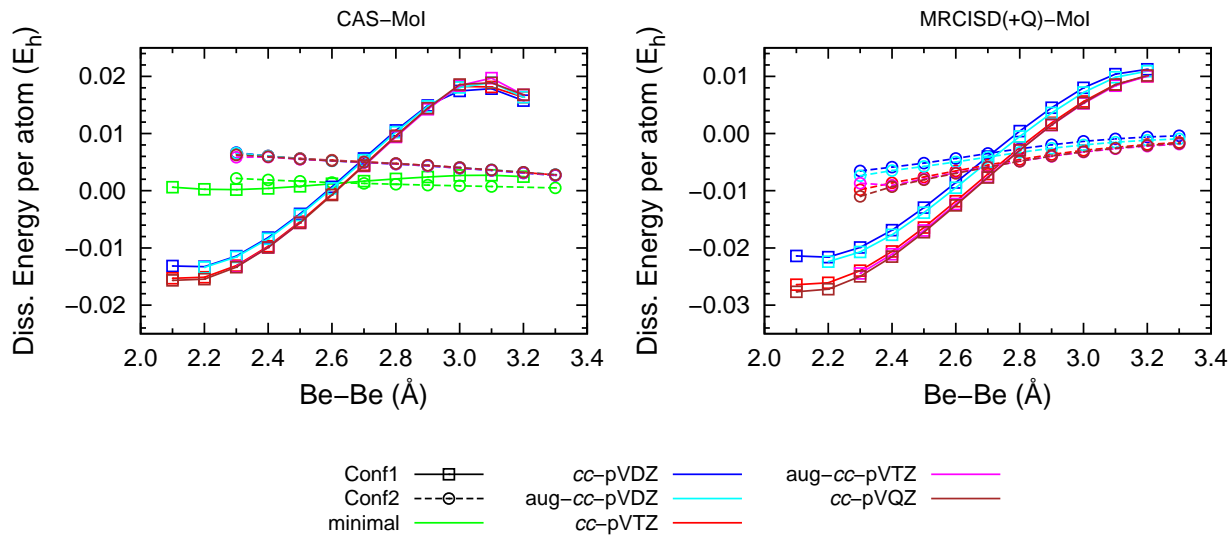}
\caption{(Color online) Dissociation energies per atom for the Be$_6$ ring obtained with CAS- and ACPF-MoI by employing different basis sets, including a minimal one, $cc$-pVDZ, $cc$-pVTZ, aug-$cc$-pVDZ, aug-$cc$-pVTZ and $cc$-pVQZ. The two starting configurations \textit{Conf1} and \textit{Conf2} were used and up to three-body increments were included. The energies are presented with respect to the energy of an isolated Be atom on the corresponding level of theory and basis set.}
\label{fig_9}
\end{figure*}

In Fig.~\ref{fig_8} we report the individual incremental contributions for both configurations. As expected, the values of the increments decrease as their order rises respecting the desired convergence expressed by Eq.~\ref{moi6}. This happens regardless of the distance regime and it ensures that the method can be correctly applied. It should not be surprising that at the one-body level, no difference is observed for all methods including dynamical correlation, since at this level only two electrons are correlated.\\
As can be seen for \textit{Conf1}, at the one-body level roughly 90 \% of the electron correlation can be obtained by CAS-MoI. On the other hand, the contributions introduced by the multireference approaches play a more important role for the two- and three-body increments where these are even larger than the fraction of correlation energy obtained by CAS-MoI. The inclusion of excitations into the delocalized virtual space leads to an energy lowering for the one- and two-body corrections, but the three-body increments are positive. In general, the difference between MRCISD-, MRCISD(+Q)- and ACPF-MoI energies is negligible while CCSD(T)-MoI values are slightly higher at the two-body and lower at the three-body level. The amount of correlation energy introduced by MRCISD(+Q)- and ACPF-MoI is about 23~\% at 2.10~{\AA} and about 25~\% at the crossing region.\\
For \textit{Conf2}, as expected, the one-body increments converge towards the atomic correlation energy as the interatomic distance increases, while the two-, three- and four-body increments drop to zero. For this reference configuration, dynamical correlation corrections seem to be particularly important as we move towards the crossing, while the effect is less important for larger internuclear distances. For instance, at 2.70~{\AA} the fraction of the introduced correlation energy is about 18~\% for both, MRCISD(+Q)- and ACPF-MoI, and it drops to about 7~\% for the isolated atoms. As can be seen, the four-body increments are one order of magnitude smaller than the three-body and have a negligible effect on the overall energy, so that the application of the MoI for more accurate basis sets and larger rings will be limited to the three-body level.\\ 
In the following, the impact of the choice of the basis set on the correlation energies retrieved with different methods, shall be discussed, using $cc$-pVDZ, $cc$-pVTZ, $cc$-pVQZ, aug-$cc$-pVDZ and aug-$cc$-pVTZ basis sets. In Fig.~\ref{fig_9} the dissociation curves calculated at the three-body level for both configurations with CAS- and ACPF-MoI and different basis sets are shown. As previously discussed, there are only minor differences between the applied methods which include dynamical correlation. Therefore ACPF-MoI was chosen as a representative example. For CAS-MoI also the results with a minimal basis set are included for comparison. Those values clearly differ from the lower lying VDZ and VTZ results for almost each internuclear distance. The inclusion of more and diffuse functions has a pronounced effect on the binding energy, as well as on the position of the crossing which is shifted towards larger Be-Be distances with increasing basis set size.\\
It is not surprising that with the inclusion of excitations into the delocalized virtuals also the basis set effects get more pronounced. This is more evident for the change from VDZ to VTZ basis sets than from the non-augmented to the augmented ones. Furthermore, a clear energetic separation occurs also for the values obtained with \textit{Conf2}, which differed less on a CAS-MoI level. \\
As can be seen, the energy difference between the values obtained with the $cc$-pVTZ and $cc$-pVQZ basis sets are significantly smaller than the separation of $cc$-pVDZ and $cc$-pVTZ values, indicating a rapid convergence of the correlation energies with respect to basis set size. To take a closer look, the cubic basis set extrapolation formula proposed by Helgaker, Klopper, Koch and Noga\cite{Helgaker1997, Klopper1998} was used along with the CAS- and ACPF-MoI results to obtain  approximate values for the correlation energy at the complete basis set limit in the minimum distance regime at 2.10~{\AA}. The fitted curves are shown in the supplementary materials. With the $cc$-pVDZ basis set, CAS-MoI yields about 93~\% of the correlation energy retrievable at the complete basis set (CBS) limit with this method, while 86~\% of the total correlation at the CBS limit is obtained with ACPF-MoI. Using the VTZ basis set, this values increase to about 97~\% and 95~\%, respectively, emphasizing the already stated accuracy of the $cc$-pVTZ basis set.
 
\subsection{Larger rings}

\begin{figure*}[ht]
CAS-MoI\\
\includegraphics[width=0.8\textwidth]{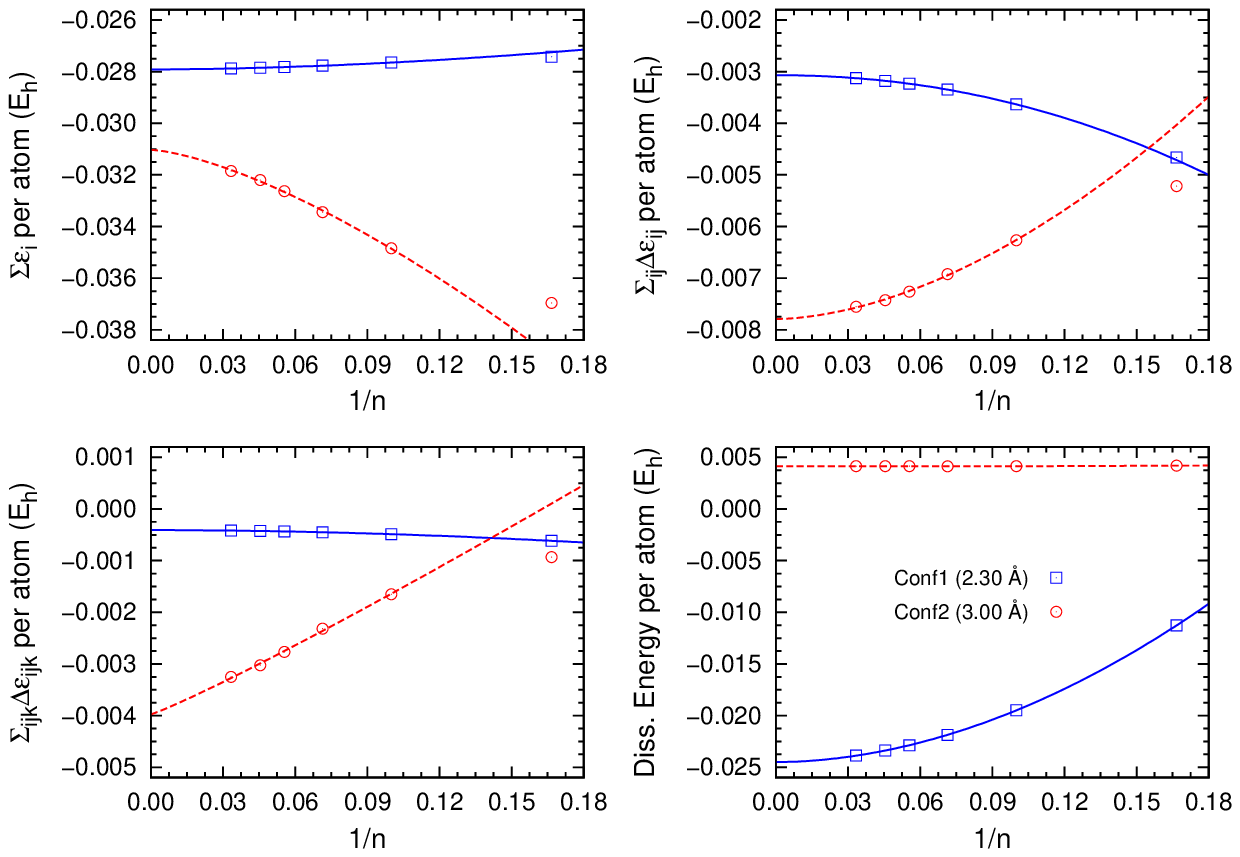}\\
ACPF-MoI\\
\includegraphics[width=0.8\textwidth]{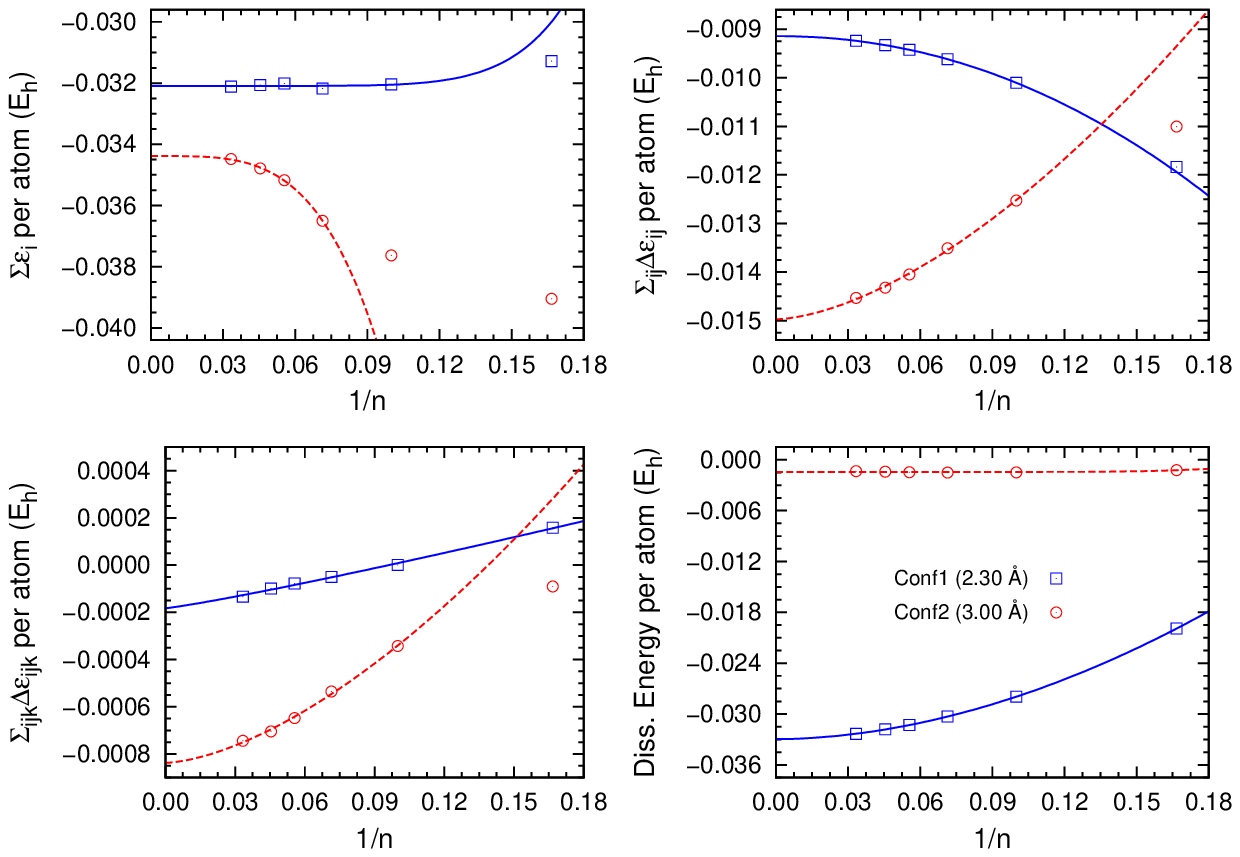}
\caption{(Color online) Correlation energy contributions of the one-, two- and three-body increments as well as the overall dissociation energy of larger Be$_n$ rings with $n=6,10,14,18,22,30$. The reported values were calculated for \textit{Conf1} at 2.30~{\AA} and \textit{Conf2} and 3.00~\AA by using CAS- and ACPF-MoI with the $cc$-pVDZ basis set. The data were fitted by means of a function of the form $a/n^b+c$. This way we extrapolated the correlation and dissociation energy for the infinite chain. The data corresponding to Be$_6$ and Be$_{10}$ were not included in a few fits where they deviate from the general trend.}
\label{fig_10}
\end{figure*}

\begin{figure*}[ht]
\includegraphics[width=\textwidth]{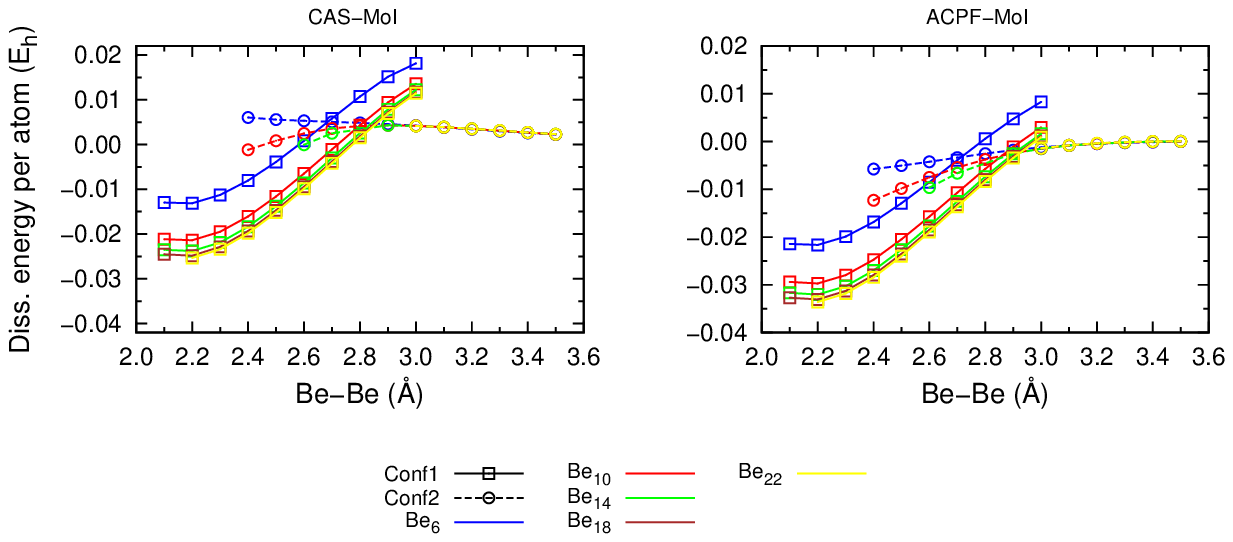}
\caption{(Color online) Dissociation curves of Be$_n$ rings, with $n=6,10,14,18,22$, calculated with CAS-MoI and ACPF-MoI by using the $cc$-pVDZ basis set. In each case, the two starting configurations \textit{Conf1} and \textit{Conf2} were employed. The energies are presented with respect to the energy of an isolated Be atom on the corresponding level of theory.}
\label{fig_11}
\end{figure*}

In larger beryllium rings, we expect a similar electronic structure to the one observed for Be$_6$ which would allow an analogous application of the MoI. Clearly the active space will be larger, but we can expect that two major configurations will play a main role in two distincted regions of the dissociation curves. In the previous investigation of Fertitta, Paulus \textit{et al.}, beryllium rings up to a size of 90 atoms were treated employing the CAS-MoI approach. By doing so, the values for the increments and the total correlation energy for the infinite system could be extrapolated.
Herein we intend to apply the same procedure for the dynamical correlation.\\
In Fig.~\ref{fig_10} we report the one-, two- and three-body increments as well as the dissociation energy per atom of Be$_n$ rings as a function of $1/n$ as calculated with CAS-MoI and ACPF-MoI. The data are reported for two internuclear distances, 2.30~{\AA} and 3.00~{\AA} and for $n=$ 6, 10, 14, 18, 22 and 30. As one can see, both correlation energy contributions present a clear trend as a function of the system size which allows us to evaluate the values corresponding to infinite chain by employing a fitting function as described in the supplementary materials. The extrapolated values are reported in Table~\ref{tab_2}. While this procedure was easily performed  for ACPF-MoI, it was not possible for other MR methods. As we have seen, for Be$_6$ the difference between MRCISD-, MRCISD(+Q)-, ACPF- and CCSD(T)-MoI is small, but this is not the case for the larger rings. The reason lies in the lack or poor correction of size-extensivity introduced by the different approaches. Indeed, since the increments are calculated as differences between correlation energies, an error is introduced if the scaling with particle number is not correctly taken into account. Such an error will eventually become particularly large for high order increments as system size increases. This can be seen by the data reported in the supplementary materials. Among the MR methods employed, only ACPF-MoI performed well, while for MRCISD- and MRCISD(+Q)-MoI show a divergence of the three-body increments. The size-extensive CCSD(T)-MoI also show a correct behavior. Finally, the dissociation curves calculated for the beryllium rings up to Be$_{22}$ with CAS- and ACPF-MoI are reported in Fig.~\ref{fig_11}. Here we can see how the dissociation curves change with system size, converging towards the limit of the infinite chain.

\begin{table*}
\centering
\caption{Extrapolated correlation energy contribution of of the one-, two- and three-body increments as well as the dissociation energy for the infinite beryllium chain. The data used for the extrapolation were obtained for Be$_n$ rings with $n=6,10,14,18,22,30$ by means of CAS- and ACPF-MoI using the $cc$-pVDZ basis set. The reported data were calculated for \textit{Conf1} at 2.30~{\AA} and \textit{Conf2} and 3.00~\AA. All values are in m$E_h$.}\label{tab_2}
\begin{tabular}{lp{0.5cm}ccp{0.5cm}cc}
    && \multicolumn{2}{c}{\textit{Conf1} 2.30~\AA} && \multicolumn{2}{c}{\textit{Conf2} 3.00~\AA} \\
          \cline{3-4}\cline{6-7}
	                       && CAS-MoI   &   ACPF-MoI && CAS-MoI   &   ACPF-MoI\\
	  \hline
	  $\sum \epsilon_{\rm i}$         &&  -27.913(1)   &  -32.09(5)    &&  -31.0(2)       &  -34.38(6)\\
	  $\sum \Delta\epsilon_{\rm ij}$  &&  -3.069(1)    &  -9.144(2)    &&  -7.8(1)        &  -15.0(2) \\
	  $\sum \Delta\epsilon_{\rm ijk}$ &&  -0.407(1)    &  -0.18(2)     &&  -3.9(2)        & -0.84(4)\\ 
\hline
	  Diss. Energy             &&  -24.50(1)    &  -32.96(1)    &&   4.143(3)      &  -1.45(4)
\end{tabular}
\end{table*}

\section{Conclusion}\label{sec_conclusion}

Different single-reference and multireference standard quantum chemical methods have been compared for small Be chains highlighting the role of static and dynamical correlation. Aiming to describe large systems for which standard methods are unfeasible, we employed the information deduced by analyzing the CI coefficients of these systems which helped us when applying the method of increments. The CAS- and MR-MoI formalisms were first applied to the Be$_6$ ring testing different approaches and the effect of the basis set. Having analyzed the CI coefficients of this system we found that two main configurations play a major role in different regions of the dissociation curve and they must be used separately in the MoI procedure as starting configuration for the localization. Assuming a similar structure of the wave function for larger rings, we applied the method to the calculation of the dissociation curve of rings up to Be$_{22}$ and performed single-point calculations for Be$_{30}$ as well. Beside the interest in describing the change of the dissociation energy with system size which allows to extrapolate values for the infinite chain/ring, we were mostly concerned in testing different MR methods within the MoI formalism and whether the method can be used for describing the whole dissociation curve. It turns out that the choice of the correlation method is crucial for the application of the MoI for large systems. Indeed, if size-extensivity is not correctly described, the error introduced into the calculation of high order increments leads to divergent behaviors. The extrapolated correlation energies for the infinite system using the CAS- and ACPF-MoI calculations suggest that the first method takes into account about 76~\% of the electron correlation in the minimum region, while the remaining 24~\% are obtained by the multireference approach. At larger distances the ratio of correlation energy introduced by CAS- and ACPF-MoI is about 85~\% to 15~\%. The obtained data are very promising, but discontinuities occur at the crossing. Further studies will be dedicated to the solution of this problem.

\acknowledgments{This research was supported by the German Research Foundation (DFG) and the Agence Nationale de la Recherche (ANR) via the project ``Quantum-chemical investigation of the metal-insulator transition in realistic low-dimensional systems'' (action ANR-11-INTB-1009 MITLOW PA1360/6-1). The support of the Zentraleinrichtung f\"ur Datenverarbeitung (ZEDAT) at the Freie Universit\"at Berlin is gratefully acknowledged. EF thanks the support of the Max Planck Society via the International Max Planck Research School. At last DK and EF would like to thank Dr. Carsten M\"uller for the fruitful discussions and suggestions.}

\bibliography{bibliography}

\newpage
\section*{Supplementary materials}

\subsection{Minima of Be$_n$ small chains}

In the manuscript we discussed the dissociation energy of Be$_n$ with $n=2,3,4,5$ obtained via different methods. The reported data were calculated in the minima of the corresponding dissociation curves. In Table~\ref{tab_sm_1} we report the position of such minima.

\begin{table*}[h]
\centering
\caption{Position of the dissociation curves of small beryllium chains calculated at different level of theory with a $cc$-pVTZ basis set. The values are reported in \AA. The precision of the position of the minima is limited by the employed grid of 0.05~\AA. The symbol "!" indicates a repulsive dissociation curve.}\label{tab_sm_1}
\begin{tabular}{lp{0.2cm}cccp{0.2cm}ccp{0.2cm}ccc}
          &&&&&& \multicolumn{2}{c}{CAS(2$n$,2$n$)} && \multicolumn{3}{c}{CAS(2$n$,4$n$) } \\
          \cline{7-8}\cline{10-12}

	  && CISD & CCSD & CCSD(T) &&  MRCISD & MRCISD(+Q) && MRCISD   &  MRCISD(+Q) &  ACPF  \\
\hline                                                                                                                                                  
Be$_2$     && 2.40&  !   &  2.55   &&  2.50   &   2.55     &&  2.50    &  2.50       &  2.50   \\
Be$_3$     && 2.30& 2.25 &  2.20   &&  2.20   &   2.20     &&  2.20    &  2.20       &  2.20   \\
Be$_4$     && 2.20& 2.15 &  2.20   &&  2.15   &   2.15     &&  2.15    &  2.15       &  2.15    \\
Be$_5$     && 2.20& 2.15 &  2.20   &&  2.15   &   2.15     &&   --     &  --         &  --    
\end{tabular}
\end{table*}

\subsection{Restricted active space calculations on the Be$_5$ chain and ring}

\begin{figure}[h]
\includegraphics[width=0.4\textwidth]{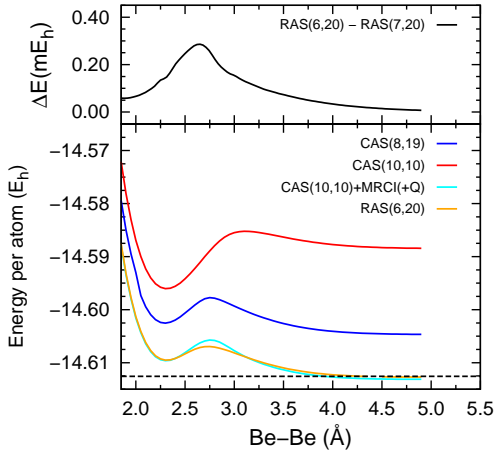}
\caption{In the lower panel, the total energy per atom of Be$_5$ linear cluster calculated with different methods and a minimal basis set. The dashed line indicates the dissociation limit calculated for a Be atom at the CAS(2,4) level. In the upper panel the difference between the ground state energy of RAS(6,20) and RAS(7,20).}
\label{fig_sm_1}
\end{figure}

As stated, Be$_5$ cannot be described at the full valence CAS(10,20) level. We took then the challenge offered by this relatively simple system to test the effectiveness of different multiconfigurational and multireference wave functions. In the manuscript we discussed RAS(6,20) results. Herein we discuss their accuracy. In Fig.~\ref{fig_sm_1} we report the potential energy curves obtained for Be$_5$ linear chain by employing the minimal basis set described in the manuscript and different approaches:
\begin{enumerate}
    \item As a first approximation, we performed CAS(8,19) calculations, by keeping the lowest lying valence orbital doubly occupied;
    \item Excluding the $\pi$ orbitals from the active space, we performed a CAS(10,10) using the $\sigma$ orbitals only;
    \item In order to improve the previous results, a MRCISD(+Q) was perfomed on top of the CAS(10,10);
    \item Finally a RAS(6,20) was performed by including all valence orbitals, but restricting the number of active electrons to six. 
\end{enumerate}	
As expected, all 20 valence orbitals are necessary for the construction of a meaningful active space able to describe dissociation achieving a size-extensive reference wave function. Indeed, at dissociation the three $p$ orbitals of each Be atom must have the same weight in the wave function. We can see how far the CAS(8,19) results are with respect to the more accurate RAS(6,20). The value at the dissociation limit of RAS(6,20) is consistent with the CAS(2,4) result obtained for a Be atom with the same basis set (dashed line in Fig.~\ref{fig_sm_1}).\\ 
It is reasonable to believe that the $\sigma$ orbitals play the major role in the construction of the bond. Indeed, the CAS(10,10)+MRCISD(+Q) results are in very good agreement with RAS(6,20) both at dissociation and at the minimum. However, the two potential energy curves deviate in the barrier region which means that in this part of the dissociation curve the role of $\sigma\rightarrow\pi$ excitations cannot be excluded.\\
Finally, we performed a RAS(7,20) calculation on Be$_5$ ring and compared it to the RAS(6,20) in order to test the accuracy of this latter method. The difference between the ground state energies for the two RAS-SCF calculations are reported in the upper panell of Fig.~\ref{fig_sm_1} and as one can see, it is lower than 0.4~m$E_h$ for any internuclear distance.

\subsection{Basis set limit extrapolation}

\begin{figure*}[h]
\includegraphics[width=0.4\textwidth]{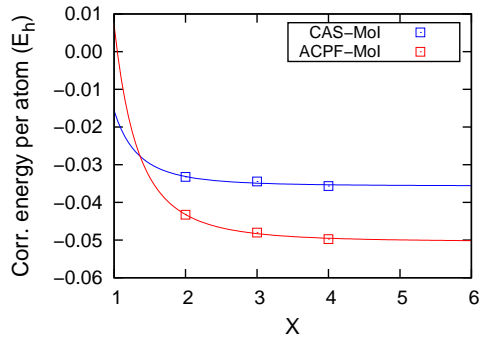}
\caption{Helgaker, Klopper, Koch and Noga extrapolation of the correlation energy per atom in the minimum distance regime at 2.10~{\AA} using CAS-MoI and ACPF-MoI. The energies are plotted against the order $X$ of the corresponding $cc$-pV$X$Z basis sets and the extrapolation converges towards the complete basis set limit of the correlation energies.}
\label{fig_sm_2}
\end{figure*}

In Fig.~\ref{fig_sm_2} we report the extrapolation to the CBS limit for the correlation energy $E_\mathrm{CBS}$ of the Be$_6$ ring calculated with CAS- and ACPF-MoI. The fit of the calculated data was performed by employing the cubic equation 
\begin{equation}
E[X] = E_\mathrm{CBS}^\mathrm{HKKN} + b X^{-3}
\label{eq_sm_1}
\end{equation}
as proposed by Helgaker, Klopper, Koch and Noga (HKKN) in a three-point extrapolation. Within this formula, $X$ denotes the cardinal numbers of the used \textit{cc}-pV\textit{X}Z basis sets, with $X = D,T,Q$ in the present case.\\
The correlation energy on each basis set level and for each method as well as the extrapolated values are listed in Table \ref{tab_sm_2}.

\begin{table*}[h]
\centering
\caption{Correlation energies obtained using CAS- and ACPF-MoI employing different $cc$-pV$X$Z basis sets with $X = D, T, Q$ and extrapolated value for the complete basis set (CBS) limit using the extrapolation scheme by Helgaker, Klopper, Koch and Noga. All values are in m$E_h$.}\label{tab_sm_2}
\begin{tabular}{lp{0.2cm}cp{0.2cm}c}
       && \multicolumn{3}{c}{$E_\mathrm{corr}$} \\
         \cline{3-5}   

	  && CAS-MoI && ACPF-MoI  \\
\hline                                                                                                                                                  
$cc$-pVDZ     && -33.227 &&  -43.286  \\
$cc$-pVTZ     && -34.454 &&  -47.997 \\
$cc$-pVQZ     && -35.680 &&  -49.751  \\ \hline
CBS limit     && -35.6(5) &&  -50.4(3)  \\
\end{tabular}
\end{table*} 


\newpage

\subsection{Extrapolation to the infinite chain}
\begin{figure*}[ht]
MRCISD-MoI\\
\includegraphics[width=0.8\textwidth]{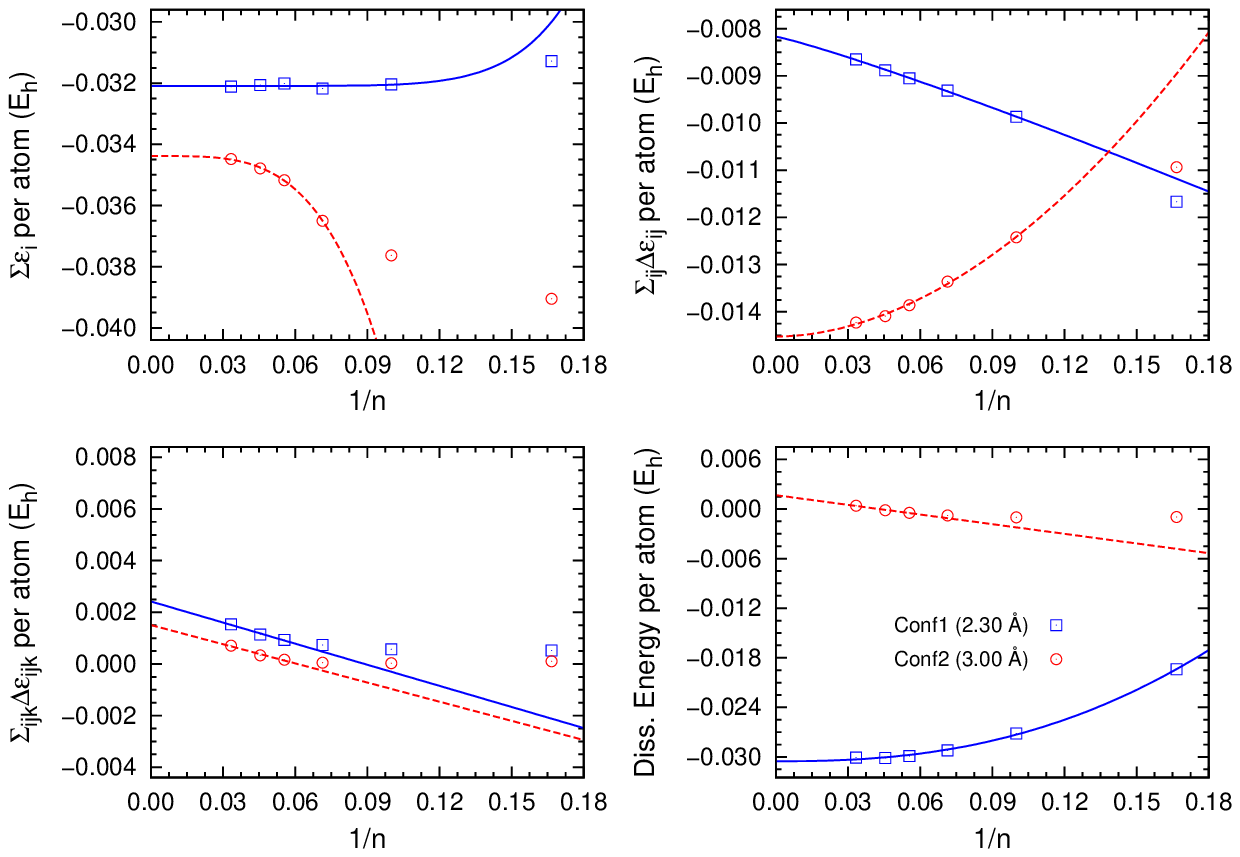}\\
MRCISD(+Q)-MoI\\
\includegraphics[width=0.8\textwidth]{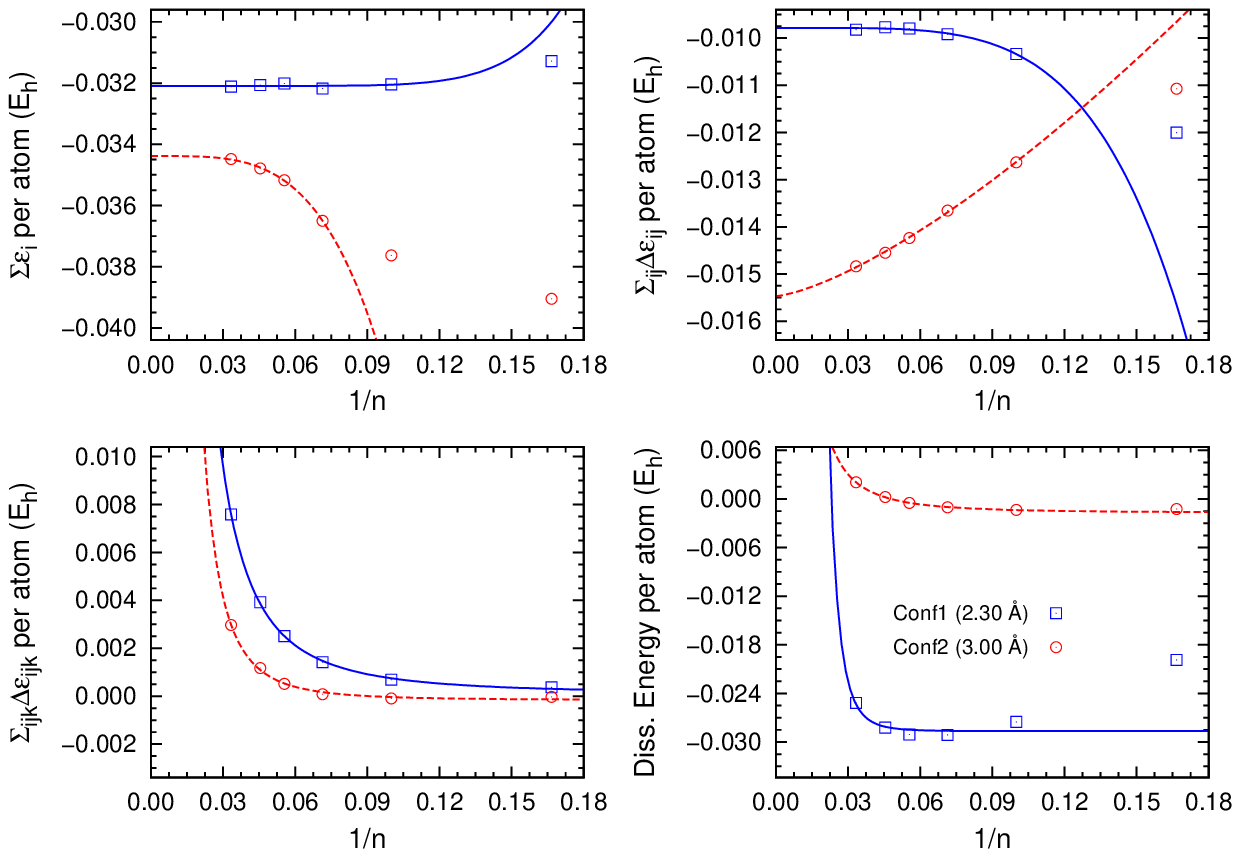}
\caption{(Color online) Correlation energy contributions of the one-, two, three-body increments as obtained with MRCISD- and MRCISD(+Q)-MoI as well as the overall dissociation energy of larger Be$_n$ rings with $n=6,10,14,18,22,30$. The reported data were calculated for \textit{Conf1} at 2.30~{\AA} and \textit{Conf2} and 3.00~\AA ~and fitted by means of a function of the form $a/n^b+c$. This way we extrapolated the correlation and dissociation energy for the infinite chain. The data corresponding to Be$_6$ and Be$_{10}$ were not included in a few fits where they deviate from the general trend.}
\label{fig_sm_3}
\end{figure*}

\begin{figure*}[ht]
CCSD(T)-MoI\\
\includegraphics[width=0.8\textwidth]{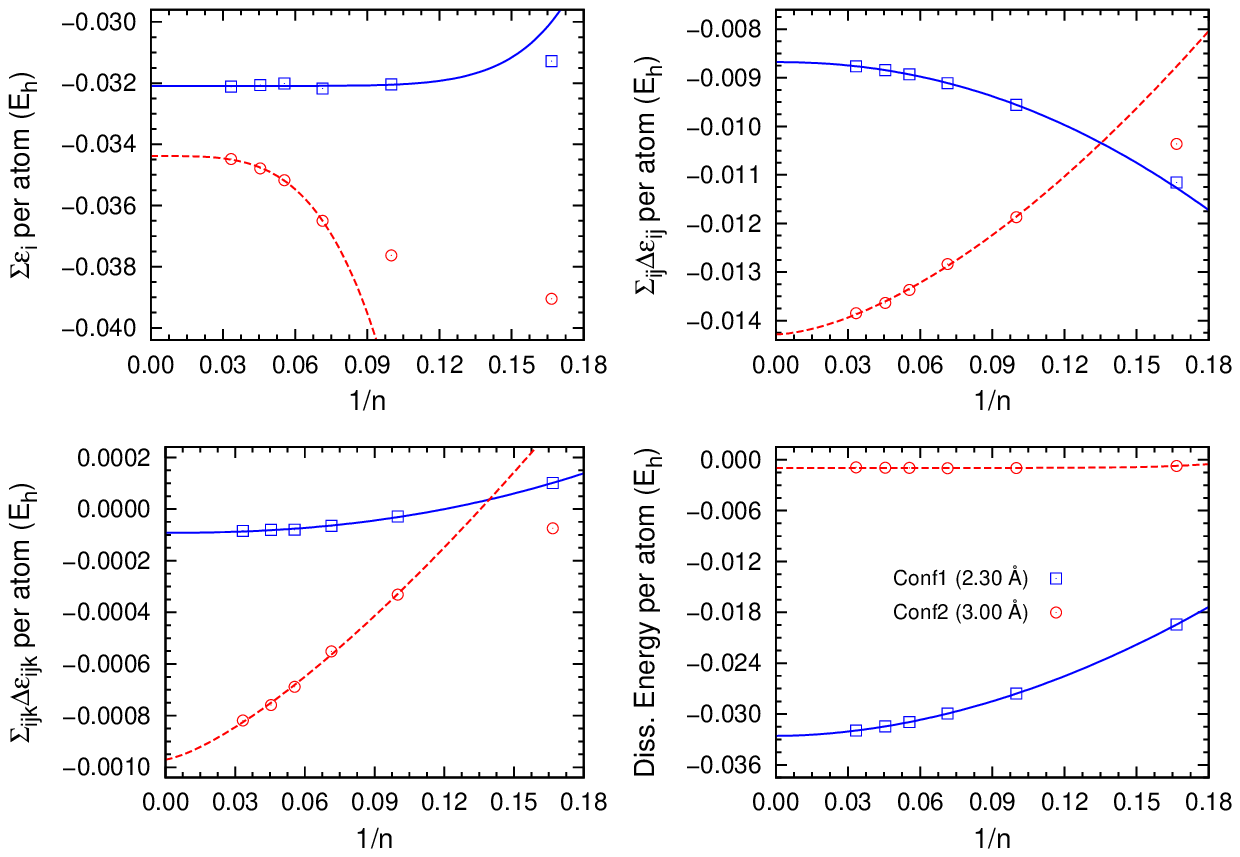}
\caption{(Color online) Correlation energy contributions of the one-, two, three-body increments as obtained with CCSD(T)-MoI as well as the overall dissociation energy of larger Be$_n$ rings with $n=6,10,14,18,22,30$. The reported data were calculated for \textit{Conf1} at 2.30~{\AA} and \textit{Conf2} and 3.00~\AA ~and fitted by means of a function of the form $a/n^b+c$. This way we extrapolated the correlation and dissociation energy for the infinite chain. The data corresponding to Be$_6$ and Be$_{10}$ were not included in a few fits where they deviate from the general trend.}
\label{fig_sm_5}
\end{figure*}

Using the MRCISD-, MRCISD(+Q)- and CCSD(T)-MoI formalisms, the sum of all incremental contributions on the one-, two- and three-body level were obtained for Be$_n$ rings with $n=6,10,14,18,22,30$, starting from the two configurations \textit{Conf1} and \textit{Conf2}. At internuclear distances of 2.30 \AA ~ and 3.00~{\AA} , these contributions to the correlation energy were extrapolated using functions of the form $a / n^b + c$ to approximate the values for an infinite chain of Be atoms, as shown in Figs. \ref{fig_sm_3} and \ref{fig_sm_5}. Besides the correlation energy contributions, also the dissociation energy was extrapolated in an analogous fashion. The resulting values for the different methods at the two distances are listed in Table \ref{tab_sm_3}.

\begin{table*}[h]
\centering
\caption{Extrapolated correlation energy contributions of the one-, two- and three-body increments as well as the dissociation energy for the infinite beryllium chain. The data used for the extrapolation were obtained by means of MRCISD-, MRCISD(+Q)- and CCSD(T)-MoI for Be$_n$ rings with $n=6,10,14,18,22,30$. The reported data were calculated for \textit{Conf1} at 2.30~{\AA} and \textit{Conf2} and 3.00~\AA. The bracketed values indicate the absolute error on the last given digit. All values are in m$E_h$.}\label{tab_sm_3}
\begin{tabular}{lp{0.1cm}p{1.9cm}p{2.7cm}p{1.9cm}p{0.1cm}p{1.8cm}p{2.8cm}p{1.8cm}}
    && \multicolumn{3}{c}{\textit{Conf1} 2.30~\AA} && \multicolumn{3}{c}{\textit{Conf2} 3.00~\AA} \\
          \cline{3-5}\cline{7-9}
        
	                       && \small{MRCISD-MoI} & \small{MRCISD(+Q)-MoI}  &  \small{CCSD(T)-MoI} && \small{MRCISD-MoI} & \small{MRCISD(+Q)-MoI}  &  \small{CCSD(T)-MoI}\\
	                
	  \hline
	  $\sum \epsilon_{\rm i}$         &&  -32.09(6) &  \centering{-32.09(6)}  &   -32.09(6)   &&  -34.38(6)  &   \centering{-34.38(6)}  & -34.38(6) \\
	  $\sum \Delta\epsilon_{\rm ij}$  &&  -8.1(1) &   \centering{-9.784(8)}  &  -8.677(1)    &&   -14.5(1)   &  \centering{-15.5(2)}  &  -14.3(1) \\
	  $\sum \Delta\epsilon_{\rm ijk}$ &&  ~2.4(2) &  \centering{$\rightarrow +\infty$}   &  -0.092(3)     &&  ~1.5(1)    &  \centering{$\rightarrow +\infty$}  & -0.97(6)\\ 
	  \hline
	  Diss. Energy             &&  -30.5(2)  & \centering{$\rightarrow +\infty$} &  -32.574(3)    &&  ~1.7(2)   &  \centering{$\rightarrow +\infty$}  &  -0.95(2)
\end{tabular}
\end{table*}


\newpage
\subsection{Individual increments on a two- and three-body level}

\begin{figure*}[ht]
Be$_6$\\
\includegraphics[width=0.8\textwidth]{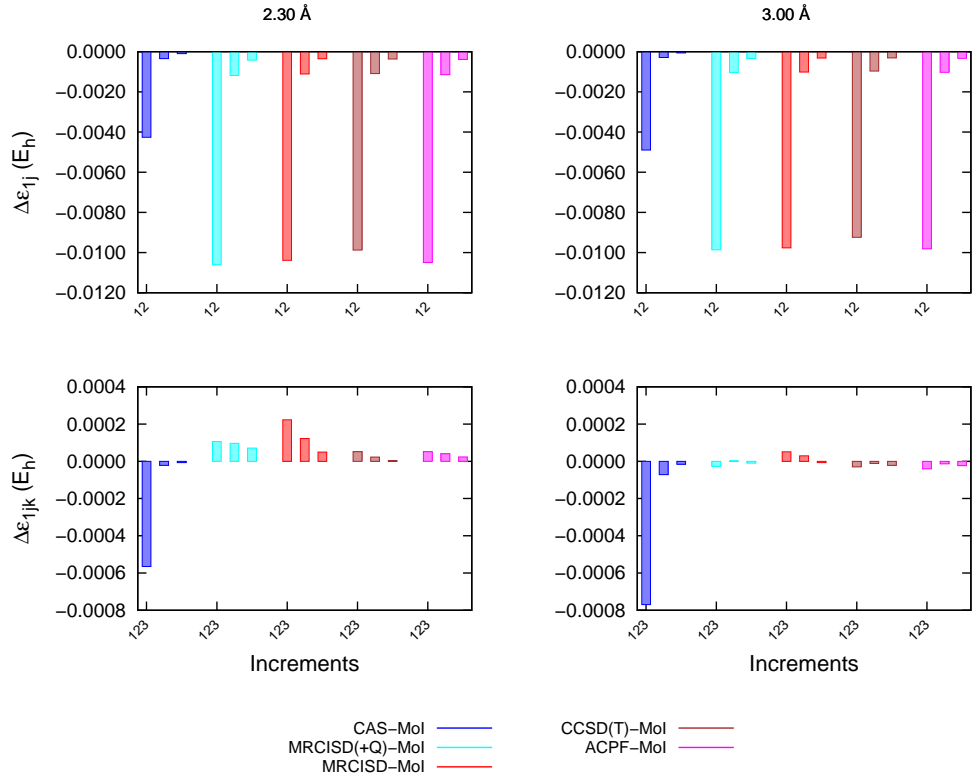}
\caption{(Color online) Correlation energy contributions of the individual two- and three-body increments $\Delta \varepsilon_\mathrm{1j}$ and $\Delta \varepsilon_\mathrm{1jk}$ as obtained with CAS-, ACPF-, MRCISD- and MRCISD(+Q)- and CCSD(T)-MoI of the Be$_\mathrm{6}$ ring. The reported data were calculated for \textit{Conf1} at 2.30~{\AA} and \textit{Conf2} and 3.00~\AA , the increments for every method are firstly ordered by increasing index $j$ and (in the case of the three-body increments) secondly by increasing $k$.}
\label{fig_sm_6}
\end{figure*}

\begin{figure*}[h]
Be$_{18}$\\
\includegraphics[width=0.8\textwidth]{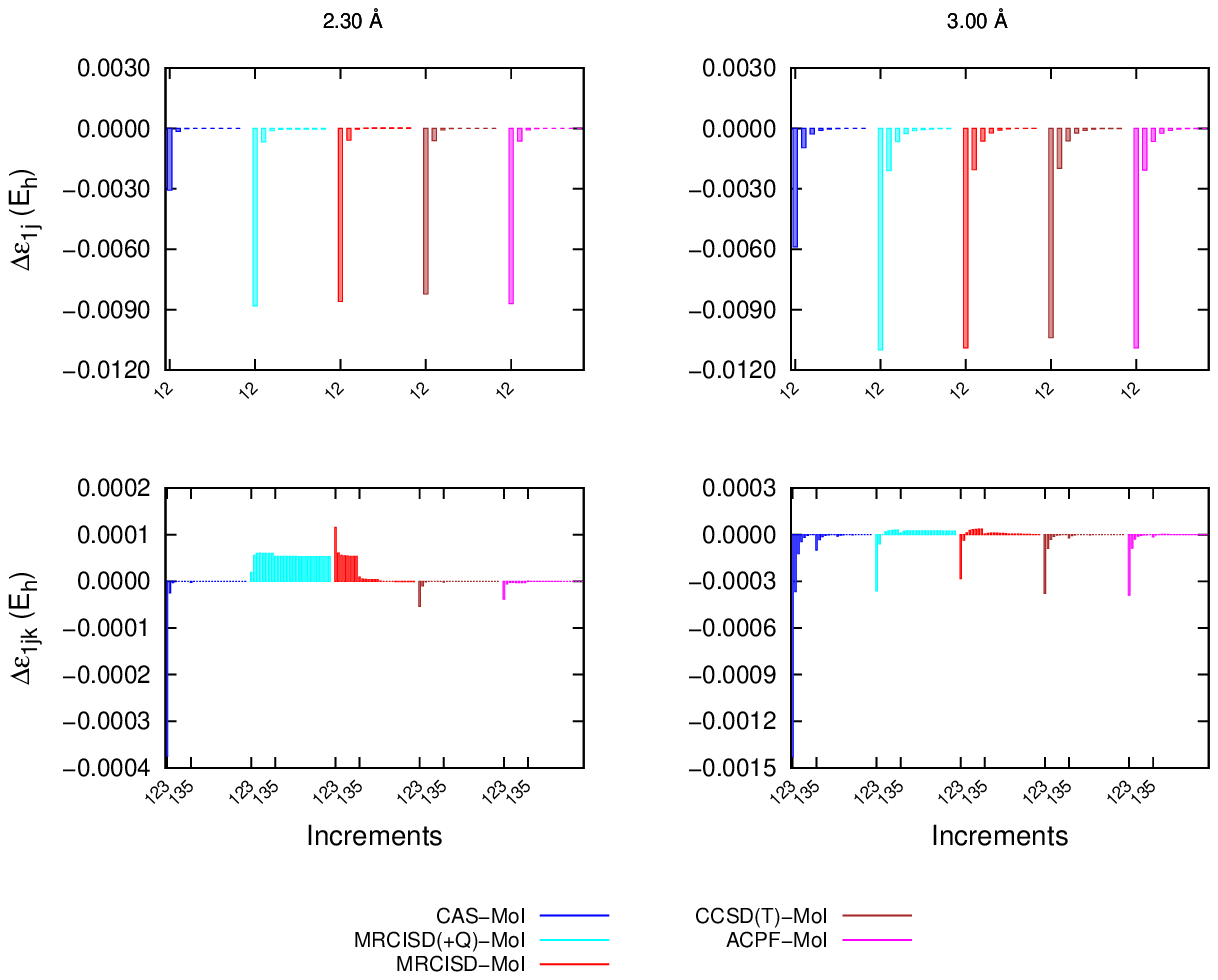}
\caption{(Color online) Correlation energy contributions of the individual two- and three-body increments $\Delta \varepsilon_\mathrm{1j}$ and $\Delta \varepsilon_\mathrm{1jk}$ as obtained with CAS-, ACPF-, MRCISD- and MRCISD(+Q)- and CCSD(T)-MoI of the Be$_\mathrm{18}$ ring. The reported data were calculated for \textit{Conf1} at 2.30~{\AA} and \textit{Conf2} and 3.00~\AA , the increments for every method are firstly ordered by increasing index $j$ and (in the case of the three-body increments) secondly by increasing $k$.}
\label{fig_sm_7}
\end{figure*}

\begin{figure*}[h]
Be$_{30}$\\
\includegraphics[width=0.8\textwidth]{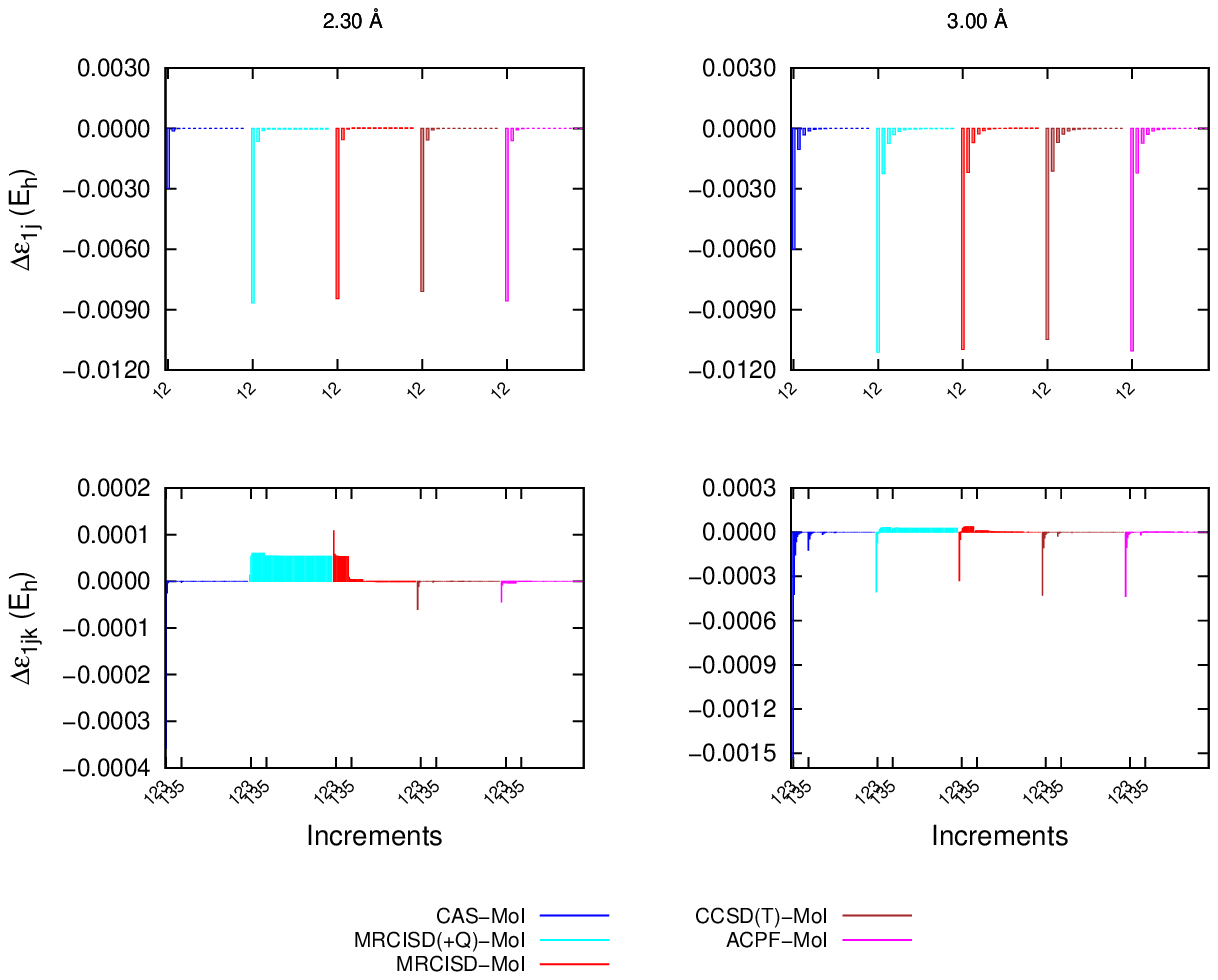}
\caption{(Color online) Correlation energy contributions of the individual two- and three-body increments $\Delta \varepsilon_\mathrm{1j}$ and $\Delta \varepsilon_\mathrm{1jk}$ as obtained with CAS-, ACPF-, MRCISD- and MRCISD(+Q)- and CCSD(T)-MoI of the Be$_\mathrm{30}$ ring. The reported data were calculated for \textit{Conf1} at 2.30~{\AA} and \textit{Conf2} and 3.00~\AA , the increments for every method are firstly ordered by increasing index $j$ and (in the case of the three-body increments) secondly by increasing $k$.}
\label{fig_sm_8}
\end{figure*}

The contributions of the individual increments at each body level obtained with the MRCISD-, MRCISD(+Q)- and CCSD(T)-MoI formalisms are depicted in Figs. \ref{fig_sm_6}, \ref{fig_sm_7} and \ref{fig_sm_8} for different sizes of the Be$_n$ rings with $n=6,18,30$. The increments are labeled by the positions of the included bodies. Because all bodies are equivalent, the first one is always fixed to $i=1$ and the following digits indicate the relative position of the second and eventually third body. The three-body increments $1jk$ are ordered by increasing $j$ and then increasing $k$, leading to the sequence $123, 124, ..., 135, 136, ...$.   
    
\end{document}